# Visualizing Streaming Text Data with Dynamic Maps

Emden R. Gansner, Yifan Hu, and Stephen North

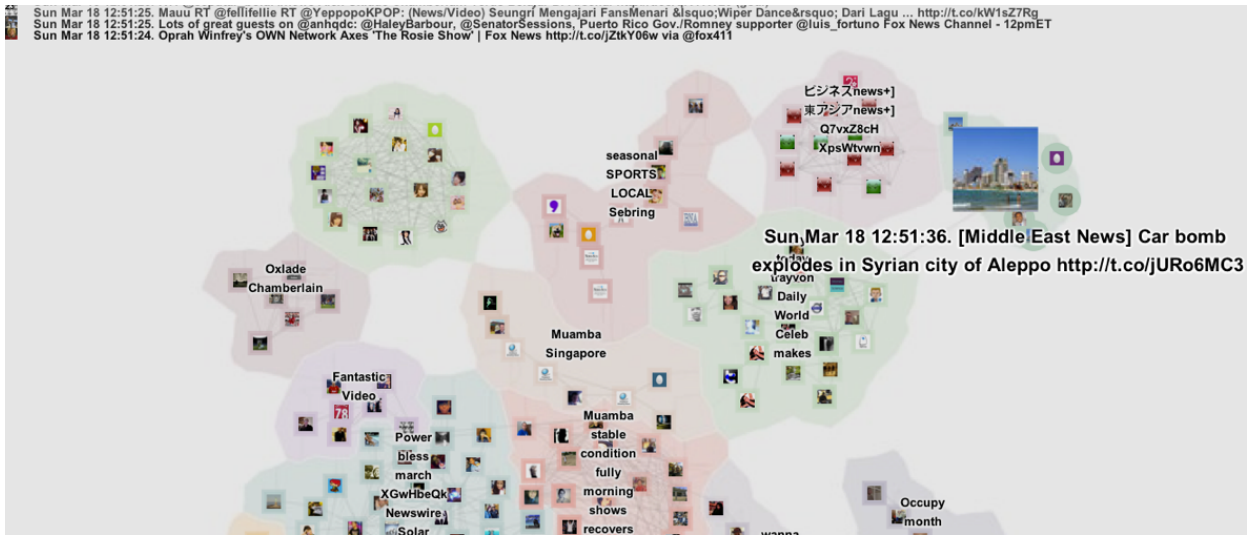

Fig. 1. TwitterScope visualization of news on March 18, 2012

**Abstract**—The many endless rivers of text now available present a serious challenge in the task of gleaning, analyzing and discovering useful information. In this paper, we describe a methodology for visualizing text streams in real time. The approach automatically groups similar messages into "countries," with keyword summaries, using semantic analysis, graph clustering and map generation techniques. It handles the need for visual stability across time by dynamic graph layout and Procrustes projection techniques, enhanced with a novel stable component packing algorithm. The result provides a continuous, succinct view of evolving topics of interest. It can be used in passive mode for overviews and situational awareness, or as an interactive data exploration tool. To make these ideas concrete, we describe their application to an online service called TwitterScope.

## 1 INTRODUCTION

With increased use of social media (Facebook, Twitter, and non-English equivalents such as Weibo), the age of big data is upon us. Some problems concerning big data are characterized solely by size, e.g., draw a Facebook friendship graph. Other data is better modeled as a stream of small packets or messages, e.g., Facebook posts, cell phone messages, or Twitter tweets. Here the problem of size relates to the rate at which these events occur. Taking Twitter as an example, one estimate gave, on average, three thousand tweets per second in February 2012, a six-fold increase over the same month two years earlier. Creating models, algorithms and tools to make sense of massive volumes of streaming data is an active area of research.

One approach is to provide a methodology for a real-time visualization and summation of the data that will provide an overview of the data and how it is changing. For this, we must take several factors into account. First, and most obvious, the velocity of the data means that it arrives in an almost continuous fashion. We need to handle this data stream efficiently, merging each packet with existing data and updating the visualization in a way that preserves the user's mental map.

Second, messages that are close in time often exhibit similar characteristics. For example, a discussion may arise in Twitter generating considerable cross-copying, or rewriting, of similar content. Our understanding can be greatly improved if related packets are appropriately grouped, categorized and summarized.

Finally, while a visual and semantic summary can provide a global view of the data stream, the user may also need tools to explore details once a topic of interest is identified.

In this paper, we propose a technique for visualizing and analyzing streaming packets. To make the presentation more concrete, we focus on streams of text packets, specifically, Twitter tweets, but it is clear how the approach can be abstracted to other types of attributed data. This approach is characterized by the following features:

- A succinct visual classification and textual summary, with details on demand. We use semantic analysis to compute similarity of tweets. Data items are arranged in space to place similar items together. Related items are discovered using a clustering algorithm, and visualized as countries using a geographic map metaphor. Countries are labeled by keywords. Details of each tweet can be revealed by rollovers, and the original source and link can be discovered by a mouse click.

- Real-time monitoring and dynamic visualization. We continuously monitor the Twitter message stream, pushing new items to the visual foreground. In doing so, our approach uses a dynamic graph layout algorithm and Procrustes projection to ensure that the layout is stable. Furthermore, when unrelated clusters of tweets form separate components, we apply a novel stable pack-

*The authors are with AT&T Labs – Research. Email:*
*{erg,yifanhu,north}@research.att.com*

ing algorithm to ensure that the visualization is stable even on a component level. Our system also allow the user to shift the display to an earlier point in time to explore what was discussed in the past.

We describe our approach as instantiated in a system called Twitter-Scope. TwitterScope runs in any HTLM5 client, and is thus accessible to a broad audience. It can be run either in passive mode for monitoring topics of interest, or as an active tool for exploring Twitter discussions on any topic. Fig. 1 shows a typical snapshot. TwitterScope is available at http://bit.ly/HA6KIR

Our system allows users to follow the data stream development in real time, observing emerging areas of interests and events. Throughout the paper we will be showing screen shots of the system. These screen shots are not a retrospective look back into the history, but rather are taken as we used the system ourselves, and illustrate how a user would use the system in discovering evolving phenomena.

The remainder of the paper is organized as follows. In the next section, we review related work on visualizing dynamic text data. This is followed, in Sections 3 , 4, by a detailed discussion of the core of our approach: the algorithms for semantic analysis, clustering and mapping, dynamic layout, and a novel algorithm for stable packing of disconnected components. We give an overview of TwitterScope in Section 5, and describe its interactive features. We conclude with a brief discussion including directions for future work in Section 6.

## 2 Related Work

Visualizing streams of text is an active area of research, especially with the growth of social media and related interest in gaining insight about this data. Šilić [29] gives a recent survey. Much of this work has a different focus from ours, dealing with the statistical analysis and presentation of topics or terms, focusing on the emergence of topic events [7], and often relying on off-line processing [25]. Our approach focuses on the visualization of the text messages themselves, arriving in real-time. Topic information emerges from the display of the messages.

For example, the TextPool visualization proposed by Albrech-Buehler *et al.* [1] scans summaries of new articles, gathers important terms from the packets, determines connectedness between terms based on co-occurrence and other measures, and draws the resulting graph. The graph evolves over time as terms and edges appear and disappear; graph stability is achieved by using force-directed placement. There is always just one component.

A context-preserving dynamic text visualization proposed by Cui *et al.* [8] displays word clouds to glean information from a collection of documents over time, keeping their layout stable using a spring embedding algorithm, aided by a mesh constructed over existing nodes.

TopicNets [21] analyzes large text corpora off-line. The main view is of a document-topic graph, though the system also allows aggregate nodes. Topic similarity is determined using Latent Dirichlet Allocation (LDA), with topics then positioned using multidimensional scaling. After topics fixed, documents are positioned using force-directed placement. The time dimension is handled by a separate visualization, with documents placed chronologically around a (broken) circle, and connected to related topic nodes placed inside the circle.

Alsakran *et al.* offer another approach in STREAMIT [2]. This system displays messages or documents positioned using a spring-based force-directed placement, where the ideal spring length is based on similarity of keyword vectors of documents. Scalability is achieved by parallelization on a GPU. Clustering is based on geometry, a second-order attribute, rather than directly on content. Clusters are labeled with the title of the most recent document joining the cluster.

Our work differs from the above in that we use a map metaphor to visualize clusters. We do not assume that the underlying content graph is connected. Our proposal supports monitoring text streams in real time.

In our approach, for semantic analysis, we considered the options of either LDA, or term frequency-inverse document frequency, to derive message-to-message similarity. We find clusters directly from this similarity. Our visualization is focused on the discrete messages as they arrive, and rather than presenting just items or node-link diagrams, we display messages within a geographic map metaphor, where topics are depicted as countries. This provides an aid to comprehension [12]. Edges are still shown to give a sense of connectivity, as there may not be edges between some topic countries. These objectives create new problems in maintaining visual stability over time. In the our approach, beyond the stasis provided by basic graph layout, we apply Procrustes transformation to documents within a component, and propose a stable packing algorithm for components to achieve further stability.

Focusing on the underlying layout algorithms, we note the large body of prior research on dynamic graph layout, in which the goal is to maintain a stable layout of a graph as it is modified via operations, typically insertion or deletion of nodes and edges. Brandes and Wagner adapt a force-directed model to dynamic graphs using a Bayesian framework [5], while Herman *et al.* [22] rely on good static layouts. Diehl and Görg [9] consider graphs in a sequence to create smoother transitions. Brandes and Corman [3] present a system for visualizing network evolution in which each modification is shown in a separate layer of a 3D representation, with nodes common to two layers represented as columns connecting the layers. Thus, mental map preservation is achieved by pre-computing good locations for the nodes and fixing those positions across the layers. Most recently, Brandes and Mader [4] studied offline dynamic graph drawings and compared various techniques such as aggregation (forming a large graph of all nodes of all time steps), anchoring (tying nodes to the previous positions) and linking (linking nodes from successive time steps, [11]), and found linking to be a generally preferred approach, followed by anchoring. While the issue of mental map preservation adopted in such approaches remains an important research issue, we note that in our application we want to avoid distortion of the layout for the purpose of mental stability. This is because such distortions introduce biases that can persist over time. In addition we want a fast procedure suitable for online dynamic layout. Hence we do not use techniques such as anchoring or linking, or ones that rely on offline processing of a given sequence of graphs.

Although there does not appear to have been much previous work on stably packing complex objects (Section 4), there has been much work on the related problem of stably removing node overlap (*e.g.* [10, 15, 19]). This has been limited to removing overlaps of simple shapes, and not aimed at packing them. On the other hand, there has also been work on optimal unrestricted placement of complex shapes. Landesberger *et al.* [31] studied ways of clustering disconnected components for visual analysis. Freivalds *et al.* [14] proposed a polyomino-based algorithm for packing disconnected components. And Goehlsdorf *et al.* [20] introduced new quality measures to evaluate two-dimensional configurations, which yield more compact layouts.

## 3 Dynamic Visualization

Our aim is to provide a visual search engine, presenting a collection of messages matching a search expression. Given a collection of messages, we would like to make an overview that helps people to make sense of it – how many messages are there? How are they related? What are the groups or clusters, which messages and authors belong to each, and what topics are discussed? How does interest in topics change over time? Finally, we would like to allow the user to drill down to the individual messages after she has identified interesting clusters, and move the display back in time to explore how a topic emerged.

### 3.1 Semantic Analysis

The first step is to cluster the messages into sub-topics using semantic analysis. One of the most effective methods for the semantic analysis of textual data is Latent Dirichlet Allocation (LDA) [**?**]. This is a generative model, which assumes that documents are distributions over topics, and topics are distributions over words. It finds these distributions by a generative process that tries to fit the observation of docu-ments over words. Once the probability distribution is known through the LDA process, so that the probability of document $i$ in topic $k$ is $p_{i,k}$, the dissimilarity of two documents $i$ and $j$ can be calculated using the Kullback-Leibler (KL) divergence

$$D_{KL}(i,j) = \sum_k p_{i,k} \ln \frac{p_{i,k}}{p_{j,k}}$$

Since the Kullback-Leibler divergence is not symmetric, in practice we use the symmetrized version, $D_{KL}'(i,j) = \frac{1}{2}(D_{KL}(i,j) + D_{KL}(j,i))$, and define similarity as $1/(1 + D_{KL}'(i,j))$.

A simpler and computationally cheaper alternative is to calculate similarity using word counts. Since naive use of word counts can be biased toward commonly used words, we use a scaled version based on term frequency-inverse document frequency (tf-idf). Let $D$ be the set of documents, $d \in D$ a document consisting of a sequence of words (terms), and $t$ a particular term of interest in $d$. Then the scaled word count based on tf-idf is:

$$\text{tf-idf}(t,d) = \text{tf}(t,d) * \text{idf}(t,D)$$

where $\text{tf}(t,d)$ is the fraction of times the term $t$ appears in $d$, and $\text{idf}(t,D)$ is the logarithm of the inverse of the proportion of documents containing the term $t$.

$$\text{idf}(t,D) = \ln \frac{|D|}{|d' \in D \text{ and } t \in d'|}$$

Similarity of documents can be calculated by cosine similarity of the tf-idf vectors.

Before computing similarity using either LDA or tf-idf, we first clean the text of the messages, removing markup or tag notations and leaving only plain content-carrying words. For example, with Twitter data, we remove terms often seen in tweets that are not important in computing similarity, such as "RT" (retweet). Another common term appearing in many tweets are "mentions," such as "@joe." Sometimes multiples of these tags appear in one tweet. We filter out these terms since they may contain unique words which would rank highly with tf-idf, but are actually not very important to matching the content. In addition, we filter out URLs which, in tweets, are almost always in a shortened alias form, and are not necessarily a bijection. Finally, we remove all stop words (e.g., "the" and "are"), before computing LDA or tf-idf similarity.

Comparing LDA and tf-idf, the former is sophisticated, and potentially more accurate. But in our application, we found LDA not any better than td-idf for identifying meaningful clusters. Sometimes LDA clusters messages that have no words in common, because LDA treats them as belonging to the same topic. The problem is that with short messages such as tweets, these assignments are not always meaningful. With tf-idf, messages in the same cluster must share at least some words, making the cluster easier to interpret, even if it is not always semantically "correct." For example, we observed that tf-idf assigned high similarity to the tweets, "Everything I know comes through visualization. Third eye senses" and "Many eyes: create visualization," among a data set of many tweets containing the word "visualization," because only those two tweets also contained the word "eye." Overall, like other researchers, we found that semantic analysis of very short text packets, such as tweets, is a challenging task [24], and will likely be studied for years to come.

After the similarity between messages is calculated, we apply a threshold (default 0.2) to the similarity matrix to ignore links between messages with low similarity. While this is not essential to the remaining steps described in this section, we found that applying this threshold can turn completely unrelated clusters into disconnected components in the graph of messages. This helps in making each of the clusters more rounded, and the map representation more aesthetic, but also introduces the complication of packing components in a stable manner, on top of the need for stability within components. We will discuss these issues in the rest of this and the following section.

## 3.2 Clustering and Mapping

After the similarities between messages are calculated, we treat the collection of messages as a graph – each message is a node, and every pair of messages is connected by an edge whose strength is given by the similarity between the two messages. A strength of zero corresponds to no edge being present. We draw the graph by multidimensional scaling (MDS), so that similar messages are placed close to each other. For visualization, each node has an associated glyph or icon, such as thumbnail pictures for tweets. We then apply the PRISM algorithm [15], which removes node overlaps while maintaining their relative proximity in the input layout.

We limit the number of messages displayed, depending on the available display area. Typically in a browser we show up to 500 tweets. Because we filter out similarities below a threshold, the graph may be very sparse or disconnected. We therefore also filter out singleton components.

We apply modularity clustering [26] to identify clusters of messages, using edge weights proportional to message similarities. After finding clusters, one possibility is to highlight them by assigning colors. We go one step further, and use a map metaphor [12] to depict clusters: messages are treated as towns and cities, where similar messages are close. Clusters of highly related messages then form countries, each enclosed by a boundary, and each country is assigned a color as different as possible from its neighbors within a defined palette [16].

As an example, in Fig. 1, we see part of a map, a screen shot of TwitterScope taken on Sunday March 18, 2012, related to the keyword "news." Each "city" represents a tweet, and is shown as a thumbnail picture of the person who posted the tweet. There is a small country on the right of the figure relating to news events about Syria, with one tweet highlighted. There is also a country with a Japanese keyword summary to the west. To the south of the Japanese topic is one with the keyword "Trayvon," relating to the tragedy of February 26 in which a 17-year-old unarmed African-American was shot by neighborhood watcher George Zimmerman. Three weeks later, the alleged shooter remained free, and this caused much discussion, and would soon make national news, as discussed further in Section 5. A small gray country in the southeast of the map with summary "Occupy month" contain tweets about the 6-month anniversary of the "Occupy Wall Street" movement on that day. There are also two countries connected by edges in the center with summary words "Muamba." They are related to the incident in which the English football player Fabrice Muamba suffered a cardiac arrest during a FA Cup quarter-final in London. This event took place on the previous evening, therefore there are many tweets about it. (Messages on these topics continued to form large clusters for several more days.) Thus, at a glance, we immediately see the themes of discussions in Twitter space about "news." Using a map metaphor makes the visualization seem more appealing than a plain graph layout, and clusters are easier to identify (Fig. 2).

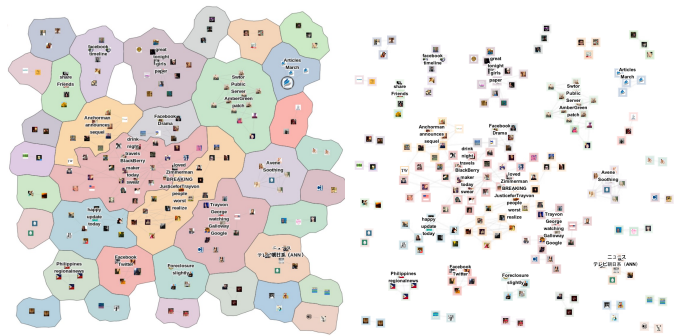

Fig. 2. A map metaphor visualization (left) seems more appealing than a plain graph layout (right), and clusters seem easier to identify.

### 3.3 Dynamic Layout

The above steps describe a static layout method. Our goal, however, is to handle dynamic text message streams, with the visualization updated frequently, say, every minute. To put this in context, consider the rate of messages in a Twitter stream. The rate clearly will depend on the initial filtering. We found that, for example, there were on average around 400 new tweets containing the keyword "news" each minute, sometimes spiking to 1200 tweets. On the other hand, for the keyword "tech," there were about 40 tweets per minute, and for "visualization," only about one new tweet per minute. Since we limit the number of messages shown, active streams, such as "news" in Twitter, require the complete replacement of every tweet by new ones at each update, and often the clusters are about completely different topics. In these situations, it may not be as important (or possible) to maintain a stable mental map. In smaller streams, for example, "visualization," only a portion of the messages will be replaced, so it is important for the remaining messages to appear at approximately the same positions, to preserve a user's mental map.

The problem of laying out dynamic graphs to preserve a user's mental map has been studied previously. One approach is to "anchor" some, or all, of the nodes, or to limit node movement by adding artificial edges linking graphs across successive time frames [11]. However, such approaches can introduce biases not in the underlying data, and that persist over time.

In our approach, we preserve the mental map via a two-step approach. Before laying out a new graph, we initialize each node in the new graph with its previous position, if it had one, otherwise at the average of its neighbors in the previous graph, and then apply MDS by local optimization. We found that MDS tends to be quite stable in terms of maintaining relative node positions, though the graph may rotate, or expand or shrink. We therefore apply Procrustes transformation [28] to the node coordinates after MDS layout. Procrustes transformation is a well-known technique [6], which attempts to find the best alignment of two inputs via rotation, translation, and scaling. Let the layout of the previous frame be $y_i$, $i = 1, 2, \ldots, |V|$, and the current layout be $x_i$, $i = 1, 2, \ldots, |V|$. If the node sets in the two layouts are different, we use the coordinates for common nodes to derive the transformation. We would like to find a translation vector $b$, scaling value $\rho$ and rotation matrix $T$ that solved the following optimization problem:

$$\text{minimize} \sum_{i=1}^{|V|} \|y_i - (\rho T x_i + b)\|^2. \tag{1}$$

The solution to this problem is

$$T = (X Y^T Y X^T)^{\frac{1}{2}} (Y X^T)^{-1}, \; \rho = \frac{tr((X Y^T Y X^T)^{\frac{1}{2}})}{tr(X X^T)},$$

where $X$ is the $2 \times |V|$ matrix of $x_i$'s. The translation vector is $b = \frac{1}{|V|} (\sum_{i=1}^{|V|} y_i - \rho T (\sum_{i=1}^{|V|} x_i))$. The minimal value of (1),

$$1 - \frac{tr((X^T Y Y^T X)^{\frac{1}{2}})}{tr(X^T X) tr(Y^T Y)}$$

is known as the Procrustes statistic.

In our application, we found that when the common set of nodes are close in the first layout, but are further apart in the second one (due to the insertion of other nodes between them), Procrustes transformation tends to scale down the new layout. In the context of abstract graph layout where nodes are represented as points, this is not a problem. However, in our setting, nodes are represented by figures having a fixed width and height, so shrinking the layout can cause node overlaps. For that reason, we limit the transformation to rotations and translations, and set $\rho = 1$.

## 4 AN ALGORITHM FOR STABLE PACKING OF DISCONNECTED COMPONENTS

The dynamic layout algorithm described in the previous section ensures that the positions of nodes are stable in successive time frames as long as the graph is connected. On the other hand, when there are multiple disconnected components the situation is more complex. While the dynamic layout algorithm aligns drawings of the same component in consecutive time frames, this alignment does not consider potential overlaps between components.

A standard technique for arranging disconnected components is to pack them to minimize total area and avoid wasted space, while avoiding overlaps. This problem has been studied in graph drawing [14, 20], in VLSI design, and in space planning for building and manufacture, where it is known as the floor-planning problem [30]. In these cases the emphasis is on maximal utilization of space. A widely used algorithm for packing disconnected graph components is that of Freivalds *et al.* [14]. It represents each disconnected component by a collection of polyominoes (tiles made of squares joined at edges.) The polyominoes are generated to cover node bounding boxes, as well as graph edges. The algorithm places components one at a time, from large to small. It attempts to place the center of polyominoes at consecutive discrete grid locations in increasing distance from the origin, until it finds one that does not overlap any components already placed. The algorithm yields packings that tend to utilize space efficiently, even though the computational complexity scales quadratically with the number of components and average number of polyominoes per component.

In the context of dynamic graph visualization, no graph packing algorithms that we are aware of were designed with maintaining layout stability in mind. For example, Fig. 3 (a) shows six disconnected components, some of them overlapping each other. If we feed this graph to a standard packing algorithm (in this case, `gvpack` in `Graphviz` [18]), we obtain Fig. 3 (b). The proximity relations are completely lost. For example, the orange component, which was below the large green component, is now to the upper right of it.

In this section we describe a packing algorithm that attempts to maintain layout stability. The algorithm extends the label overlap removal algorithm of Gansner and Hu [15], by making it work on graph components of arbitrary shape, not just rectangular node labels. In addition, the new algorithm also attempts to remove "underlap" – unnecessary space between components; thus it is a packing algorithm, not only an overlap removal method.

We begin by adapting the polyomino algorithm for component packing. For each component, polyominoes are used to cover nodes and its labels, as well as all the edges. Fig. 4 (a) shows the overlapped components of Fig. 3 (a), now covered by polyominoes.

We use the polyominoes to help detect collisions between components. We set up an underlying mesh (not shown). The mesh cell size may or may not be the same as the polyomino size. When polyominoes move due to adjustment to the position of the corresponding components, we hash the mesh cells that intersect with the polyominoes of one component, and detect a collision of this component with another component by checking the array hash table against polyominoes of the second component.

To achieve our goal of removing overlaps and underlaps, while preserving the proximity relations of the initial component configuration, we first set up a rigid "scaffolding" structure so that while components can move around, their relative positions are maintained as much as possible. This scaffolding is constructed using an approximate *proximity graph* that is rigid, in the form of a Delaunay triangulation (DT). The triangulation is carried out using the center of each component. Fig. 4 (a) shows the triangulations as thick dark lines, connecting the centers of each component.

Once we find a DT, we search every edge in it and test for component overlaps or underlaps along that edge. If there is an overlap, we compute how far the components should move apart along the direction of the edge to remove the overlap. If there is an underlap, we compute how far the components should move toward each other along the edge to close the gap. The distance of the move can be calcu-lated using an iterative algorithm, where two overlapping components are first separated enough so that they do not overlap, then a simple bisection algorithm is applied until a sufficiently accurate measure of the distance of travel is achieved. In an implementation, though, we can use a cheaper though less accurate measure. Suppose for one of these edges, the two endpoint components are $i$ and $j$. Let the edge be $\{i, j\}$, and the coordinates of the two end points be $x_i^0$ and $x_j^0$. We project the corners of each polymino onto the line $x_i^0 \rightarrow x_j^0$. If the two components overlap, it is certain that the projection of the components will overlap. The amount of overlap on line $x_i^0 \rightarrow x_j^0$ is denoted by $\delta_{ij} > 0$. On the other hand, if the two components are separated, it is likely (but not always true) that the projection of the components are separated by a gap. We denote this underlap (negative overlap) as $\delta_{ij} < 0$. If $\delta_{ij} > 0$ but no collision is detected, we set $\delta_{ij} = 0$. We define the *ideal length factor* to be

$$t_{ij} = 1 + \frac{\delta_{ij}}{\|x_i^0 - x_j^0\|} \qquad (2)$$

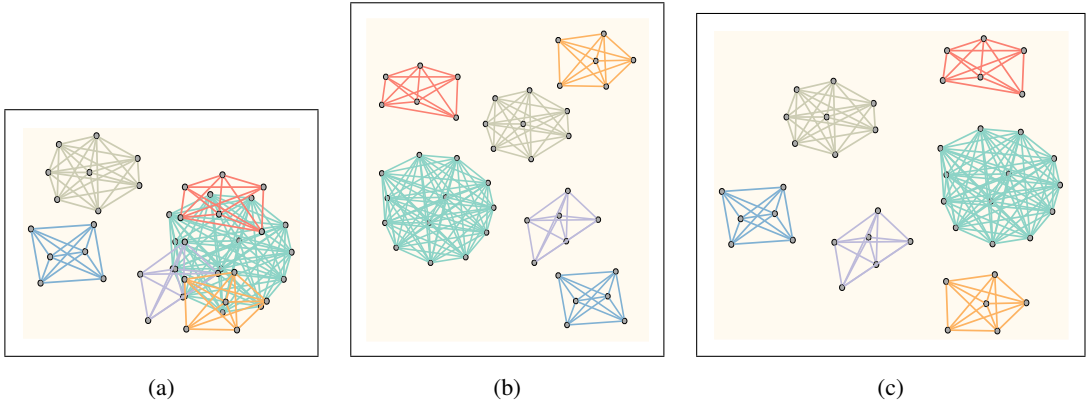

(a)            (b)            (c)

Fig. 3. Illustration of the stable packing problem. (a) Graph with overlapping components. (b) Result of a standard packing algorithm. Proximity relations among components are lost. (c) Result of proposed stable packing algorithm. Proximity information is retained.

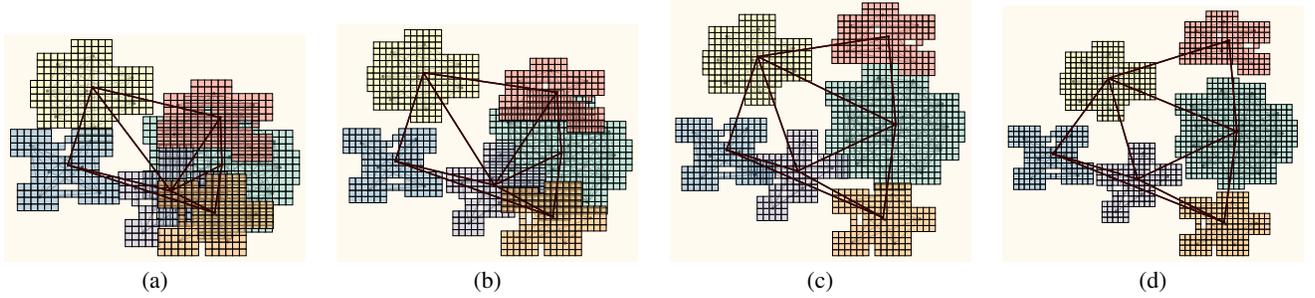

(a)        (b)        (c)        (d)

Fig. 4. Illustration of proposed stable packing algorithm. (a) Polyominoes are generated to cover nodes and edges of the components in Fig. 3. (Nodes are assumed to have nonzero width and height. To avoid clutter in the figure, they are drawn as points.) A triangulation is carried out over the centers of the components. This triangular mesh guides the overlap and underlap removal algorithm to ensure that proximity relationships are maintained. (b), (c) and (d) After 3 iterations, the overlap has been eliminated while the relative position of components remains stable. The resulting graph is shown in Fig. 3

Component overlap and underlap can be removed if we expand or shrink the edge between these components; see Fig. 5. Therefore we want to generate a layout such that an edge in the proximity graph has the ideal edge length close to $t_{ij}\|x_i^0 - x_j^0\|$. In other words, we want to minimize the following stress function

$$\sum_{(i,j) \in E_P} w_{ij} \left( \|x_i - x_j\| - d_{ij} \right)^2. \qquad (3)$$

Here $d_{ij} = s_{ij}\|x_i^0 - x_j^0\|$ is the ideal distance for the edge $\{i, j\}$, $s_{ij}$ is a scaling factor related to the overlap factor $t_{ij}$ (see (4)), $w_{ij} = 1/\|d_{ij}\|^2$ is a weighting factor, and $E_P$ is the set of edges of the proximity graph.

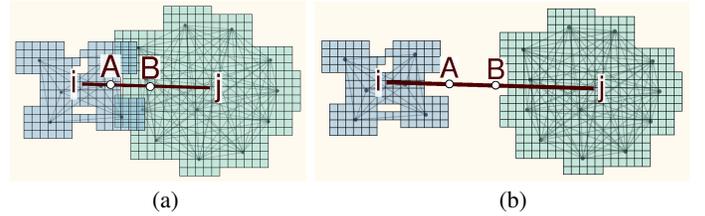

(a)            (b)

Fig. 5. Illustration of ideal length factors. (a) Components $i$ and $j$ overlap. Projection of polyominoes of $i$ on the line $i \rightarrow j$ extends as far right as $A$; projection of polyominoes of $j$ on the line $i \rightarrow j$ extends as far left as $B$. Since $|AB|$ is $1/3$ of $|ij|$, the ideal length factor is $1 + 1/3 = 1.33$. (b) Components $i$ and $j$ underlap. Since $|AB|$ is $1/4$ of $|ij|$, the ideal length factor is $1 - 1/4 = 0.75$.

Given that the DT is a planar graph with at most $3k - 6$ edges, where $k$ is the number of components, the stress function (3) has no more than $3k - 6$ terms. Furthermore, the rigidity of the triangulated graph provides a good scaffolding that constrains the relative position of the components and helps to preserve the global structure of the original layout.

Trying to removing overlaps and underlaps too quickly with the above model with $s_{ij} = t_{ij}$ can be counter-productive. Imagine the situation where components form a regular mesh configuration. Initially, the components do not overlap. Then assume that component $i$ in the center grows quickly, so that it overlaps severely with its nearby components, while the other components remain unchanged. Suppose components $i$ and $j$ form an edge in the proximity graph, and they overlap. If we try to make the length of the edge equal to $t_{ij}\|x_i^0 - x_j^0\|$, we will find that $t_{ij}$ is much larger than 1, and an optimal solution to the stress model keeps all the other vertices at or close to their current positions, but moves the large component $i$ outside the mesh entirely, to a position that avoids any overlap. This is not desirable because it destroys the relative position information. Therefore we damp the overlap and underlap factor by setting

$$s_{ij} = \max(s_{min}, \ \min(t_{ij}, s_{max})) \qquad (4)$$

and try to remove overlaps and underlaps gradually. Here $s_{max} > 1$ is a parameter limiting the amount of overlap, and $s_{min}$ a parameter limiting the amount of underlap that we are allowed to remove in one iteration. In experiments, we found that $s_{max} = 1.5$ and $s_{min} = 0.8$ work well.

After minimizing (3), we arrive at a layout that may still have component overlaps and underlaps. We then regenerate the proximity graph using DT, calculate the ideal length factor along the edges of this graph, and rerun the minimization. This defines an iterative process that terminates when there are no more overlaps or underlaps along the edges of the proximity graph. In addition, to avoid having components oscillate between overlap and underlap, we set $s_{min} = 1$ after a fixed number $N$ of iterations. In experiments, we set $N = 10$.

For many disconnected graphs, the above algorithm yields a drawing that is free of component overlaps, and utilizes the space well. For some graphs, however, especially those with components having extreme aspect ratios, overlaps may still occur, even though no overlaps are detected along the edges of the proximity graph.

To overcome this situation, once the above iterative process has converged so that no more overlaps are detected over the proximity graph edges, we hash the components one at a time in the background mesh to detect any overlaps, and augment the proximity graph with additional edges, where each edge consists of a pair of component that overlap. We then re-solve (3). This process is repeated until no more component overlaps are found. We call this algorithm proximity graph-based STAble PACKing (STAPACK) Algorithm 1 gives a detailed description of this algorithm.

---

**Algorithm 1** Proximity graph based stable packing algorithm (STAPACK)

---

Input: Disconnected components, and their centers $x_i^0$.
**repeat**
    Form a proximity graph $G_P$ of $x^0$ by Delaunay triangulation.
    Find the ideal distance factors (2) along all edges in $G_P$.
    Solve the stress model (3) for $x$. Set $x^0 = x$.
**until** (no more overlaps along edges of $G_P$)
**repeat**
    Form a proximity graph $G_P$ of $x^0$ by Delaunay triangulation.
    Find all component overlaps using background mesh hashing.
    Augment $G_P$ with edges from component pairs that overlap.
    Find the ideal length factor (2) along all edges of $G_P$.
    Solve the stress model (3) for $x$. Set $x^0 = x$.
**until** (no more overlaps)

---

We now discuss the computational complexity of the proposed algorithm. Let there be a total of $P$ polyomino cells for all graphs, and $k$ disconnected components. Let the underlying mesh have dimension $m \times m$. Setting up the mesh takes time and memory proportional to $m^2$. Delaunay triangulation can be computed in $O(k \log k)$ time [13]. We used the mesh generator Triangle [27] for this purpose.

Detecting a collision of two polyominoes takes time proportional to the number of polyominoes in the components, hence the total time for the first loop of STAPACK is $O(k \log k + m^2 + P)$

Finding all the overlaps after the first loop of STAPACK is carried out takes $O(Pq)$ time, where $q$ is the number of overlaps. Because at that stage there are no more overlaps along edges of the triangulation graph, $q$ is usually very small, hence this step can be considered as taking time $O(P)$.

The stress model is solved using stress majorization [17]. The technique works by bounding (3) with a series of quadratic functions from above, and the process of finding an optimum becomes that of finding the optimum of the series of quadratic functions, which involves solving linear systems with a weighted Laplacian of the proximity graph. We solve each linear system using a preconditioned conjugate gradient algorithm. Because we use DT as our proximity graph and it has no more than $3k - 6$ edges, each iteration of the conjugate gradient algorithm takes time $O(k)$. Hence the total time to minimize one stress function is $O(ckS)$, where $S$ is the average number of stress majorization iterations, and $c$ the average number of iterations for the conjugate gradient algorithm.

Therefore, overall, Algorithm 1 takes $O(t(ckS + k \log k + m^2 + P))$ time, where $t$ is the total number of iterations in the two main loops in Algorithm 1. The most expensive term is $tm^2$, relating to the underlying mesh. This value is dependent on the size of the polyominoes, because if the polyomino size is small, the underlying mesh should also be finer to be able to offer fine-grain collision detection. Typically a value of $m = 500$ is more than enough to offer fine resolution collision detection, and a smaller value such as $m = 100$ is often enough for the number of disconnected components we typically need to handle ($< 100$). Therefore the algorithm runs quickly.

Fig. 4 (b-d) shows the result of 3 iterations on the overlapping components in Fig. 4 (a). As can be seen, in this case the process converges quickly and the result in Fig. 4 (d) packs the components with no overlaps, while proximity is well-preserved.

## 5 TWITTERSCOPE SYSTEM OVERVIEW

We implemented a prototype system, called TwitterScope, that employs the visualization technique described above. Its aim is to provide a dynamic view of tweets related to a given keyword or set of keywords. Fig. 5 shows an overview of TwitterScope. It collects data using Twitter's streaming API methods. Twitter offers multiple APIs, including the Spritzer, which provides a small sample (around 1%) of all tweets, and a "filter" API which allows up to 400 keywords. Data is collected continuously and stored persistently. This supports queries over any time period after the collector started running, which is crucial for repeatable experiments.

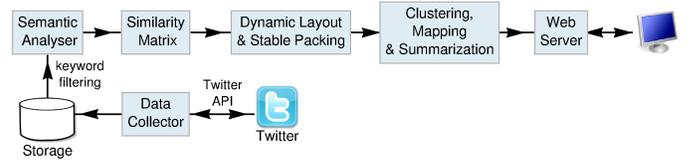

Fig. 6. TwitterScope system overview

In parallel with the data collector, a semantic analyzer finds stored tweets matching a specific keyword, and outputs a similarity matrix. By applying a threshold, we can ensure that the matrix is sparse. The graph representation of this matrix is embedded, taking into account the layout in the previous time step, using the dynamic graph layout algorithm that keeps the component layout stable across time, and also uses the stable packing algorithm when arranging disconnected components. This information is used to derive clusters, which in turn are used to produce a map of the tweets. In addition, the system produces a list of the top keywords in the tweets of a cluster which can be used to summarize the cluster. The final results are pushed to the web server, and show up on the user's browser continuously.

In certain cases, the dominance of some topics and the sparsity of tweets in others cause the layout of the graph to have an extreme aspect ratio. This is not surprising because the stable packing algorithm attempts to maintain existing proximity information. Therefore, to readjust the aspect ratio, we periodically repack afresh.

### 5.1 Passive and Active Modes of Usage

To make the system available to a broad audience, we implemented the interactive front-end in JavaScript and HTML5, that run in most web browsers. From the user's point of view, TwitterScope has two modes.

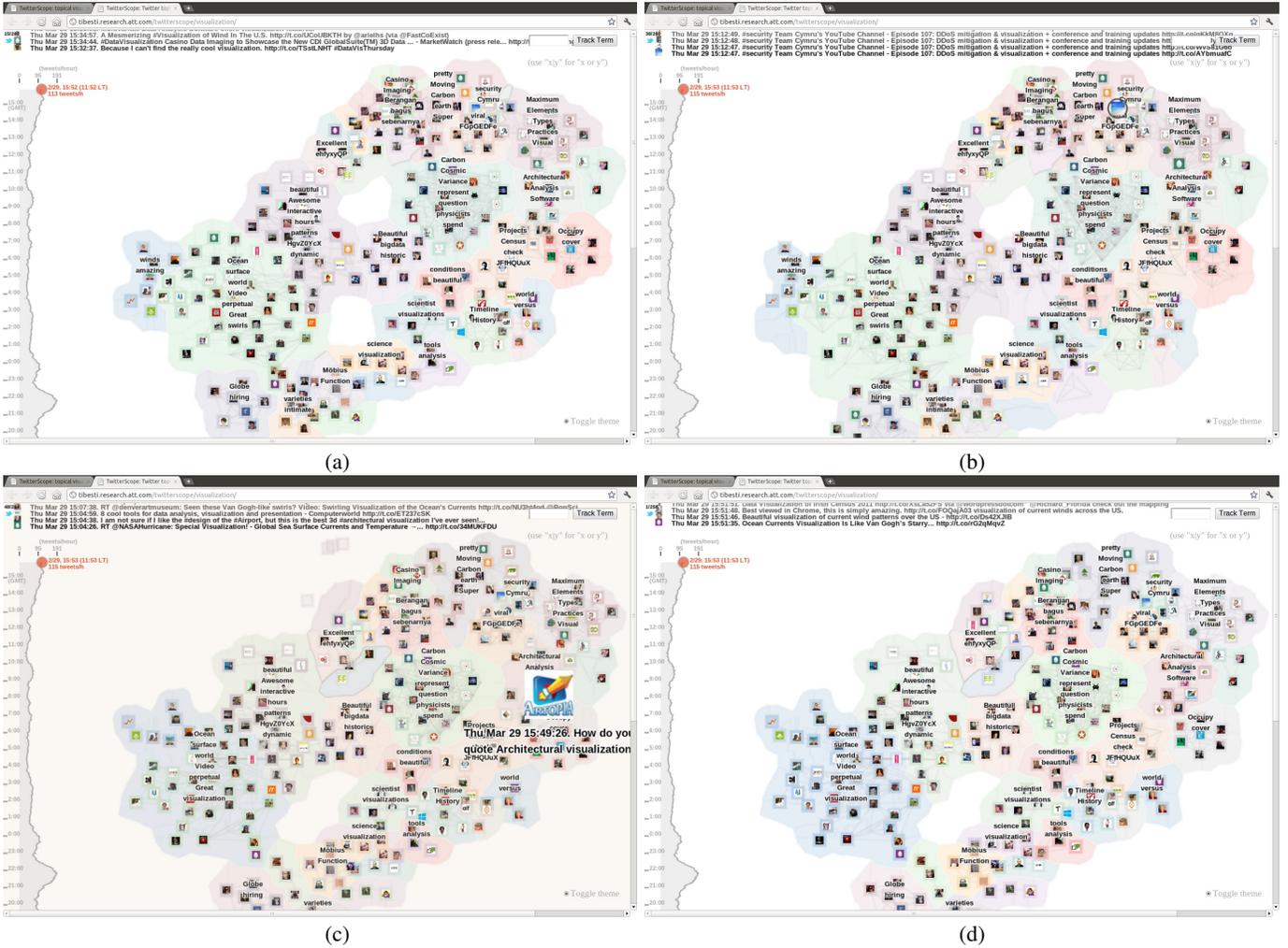

Fig. 7. Illustration of stable dynamic update. (a) A visualization of the landscape of Twitter discussion about "visualization" (11:52 local time). (b) some old tweets drop out, and positions of remaining tweets change slightly (11:53 local time). (c) four new tweets move in from top (four pink squares above the map). A tinted "veil" is applied to make the highlighted tweet more prominent, which causes a difference in background color. (d) map update finishes.

In passive mode, it runs in the background and makes overviews for monitoring the given keywords. The most recent matching tweets are shown in a rolling text panel at the top of the screen. We also highlight each tweet in turn, enlarging its thumbnail image and displaying its text. To prevent the expanded thumbnails from occluding nearby nodes, we apply fisheye-style radial distortion to move neighbors away from the highlighted node. More specifically, let the nominal size of a typical node $i$ be $w$ with position $x_i$, and the size of the highlighted node $j$ be $w'$. We want the new position for node $i$ to be

$$x_i \leftarrow x_i + \frac{a}{1+b||x_i - x_j||} \frac{x_i - x_j}{||x_i - x_j||}.$$  (5)

Thus the amount of displacement

$$f(||x_i - x_j||) = \frac{a}{1+b||x_i - x_j||},$$  (6)

decays to zero as $||x_i - x_j|| \rightarrow \infty$. If node $i$ is very close to $j$, we still want it to be visible, so we set

$$f(0) = w'/2$$  (7)

Finally we want the displacement function to decay quickly, so we set

$$f(w') = f(0)/2.$$  (8)

Substituting (7) and (8) into (6) gives $b = w'/2$ and $b = 1/w'$.

Fig. 7 shows a sequence of screen-shots illustrating the dynamic graph layout algorithm that keeps the component layout stable across time. In Fig. 7 (a), a visualization of the landscape of Twitter discussion about "visualization" is given at 11:52 local time. At 11:53, some old tweets drop out, and positions of remaining tweets changed slightly (Fig. 7 (b)). Then four new tweets move in from (four pink squares above the map, fig. 7 (c)). Finally, in Fig. 7 (d), the update finishes. Note that as dynamic update is underway, arriving tweets are continuously highlighted. For example, Fig. 7 (c) shows a highlighted tweet about architectural visualization in a country labeled "Architectural/Analysis". Notice that by applying a fisheye transformation, pictures adjacent to the highlighted one remain visible. When a tweet is highlighted, we cover the rest of the canvas with a very lightly tinted "veil" to make the highlighted tweet stand out. This accounts for the small difference in the background color of Fig. 7 (c) compared with the other three screen shots. Additional design details include use of haloes around the text of the highlighted tweet, as well as around country labels, – a feature employed in map design to make text stand out.

In active mode, the interface allows the user to interactively explore the tweet landscape. By placing the mouse pointer over the icon of a tweet, the user can see its text; clicking on the icon takes the user to the Twitter page of the person who sent the tweet, as well as to the website pointed to by a URL, if a URL link is found in the tweet. The global view displays all of the tweets associated with a set of keywords. To drill down further, the user can type a term and have the system track it. Each tweet containing the term will be indicated by a disk behind the icon, and as new tweets come in, this term will continuously be tracked and matched tweets will be highlighted.

## 5.2 Recalling and Tracking Historical Data

TwitterScope also provides an historical perspective. On the left of the display is a time series graph of the number of tweets per hour during the last 24 hours. Mousing over the time series shows the time and number of tweets, as well as a list of keywords describing the messages at that time frame; clicking on the graph moves the display to the corresponding time, allowing the user to investigate tweets that were made then.

more "bursty" nature of tweets. We make $N$ an adjustable parameter in our system, but by default set $N = 2000$, which seems to work well in capturing changes from one frame to another. Fig. 8 (top) shows keywords at 4:02 am (EST), the words "Thailand", "kill" and "Bombing" catches the user's attention. Clicking on the time frame reveals the bombing incident happened in Thailand, with a country entitled "Thailand..attacks..injured". Clicking on the highlighted tweet would take the user to a news story at the USA TODAY website about the bombing.

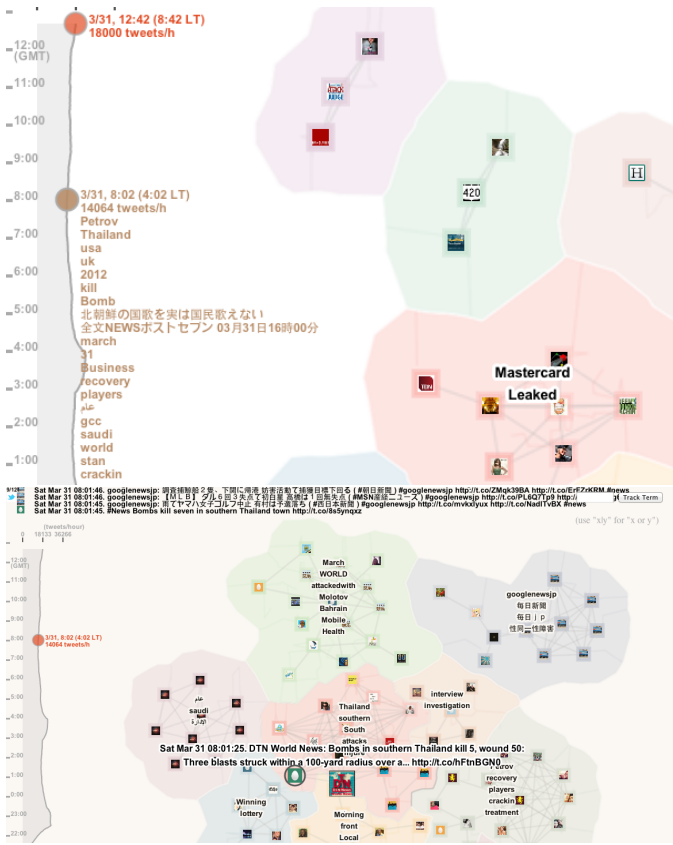

Fig. 8. Top: rolling the mouse over the time series graph reveals keywords "Thailand", "kill" and "Bombing" at the 4:02 am EST time frame. Bottom: clicking on the time frame reveals the bombing incident. Clicking on the tweet highlighted would take the user to a news story about the bombing at the USA TODAY website.

To make the keywords more meaningful and specific to a time frame, we should not just count the top words of all the tweets at that time. Doing so would cause certain commonly used words showing up frequently, make the list of top keywords uninformative. For example, words such as "good, great, sad, morning, evening" frequently appear when monitoring tweets matching the words "news". While these words can be filtered out, we take the additional measure of finding unique words, conditioning on the baseline distribution of all words in tweets. Specifically, we take a moving time window of $N$ tweets appeared prior to this time frame, merge with the new tweets, and calculate the tf-idf of all words in the tweets that appear in this time frame. We then take words that have the highest tf-idf ranking as the keywords for this time frame. A larger value of $N$ could capture keywords that appeared persistently in a large time window (e.g., if we take all tweets in the last 12 months, then the word "trayvon" could appear near the top in March 2012). A smaller value of $N$ emphasizes the

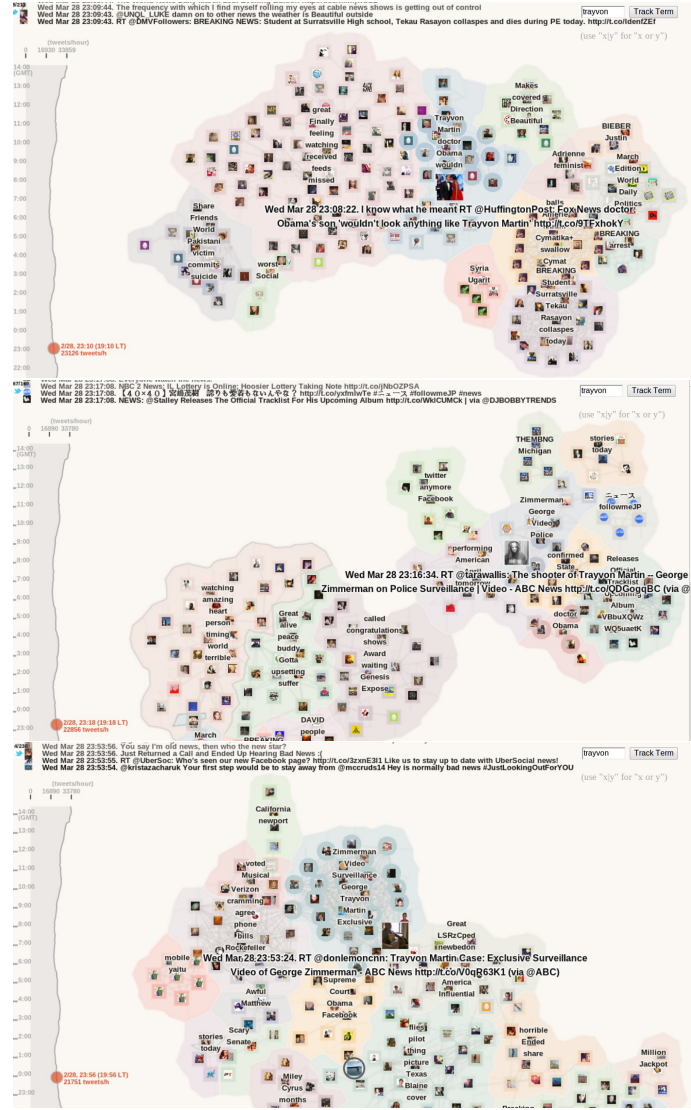

Fig. 9. An illustration of keyword tracking: the word "trayvon" is entered into the "track" search box, and we select three time frames on March 28, 2012. Top: at 19:10 there are tweets about a FoxNews item "Obama's son wouldn't look anything like Trayvon Martin"; Middle: at 19:18, tweets about a newly released police surveillance video of Zimmerman start to emerge, while tweets about the FoxNews comments decrease (pink region and disks south of the highlighted tweet); Bottom: at 19:56, tweets about the newly released video increase and become an important part of the news landscape.

If a term is being tracked prior to clicking on the time series graph, the matching tweets in that time frame will also be highlighted. Fig. 9 illustrates the use of the tracking functionality on historical data. The word "trayvon" is entered into the "track" search box, to track news on March 28, 2012 about the slain black teenager Trayvon Martin, and his killer George Zimmerman, first discussed in Section 3.2. Fig. 9 (top) shows the news landscape at 19:10 EST, there are tweets highlighted with blue disks behind the thumbnail pictures. In particular, one highlighted tweet is: "Obama's son wouldn't look anything like Trayvon Martin." This was broadcast minutes earlier on FoxNews, when one commentator said, "If the President had a son, he wouldn't look anything like Trayvon Martin. He'd be wearing a blazer from his prep school...", in reference to President Obama's comment a few days before, "If I had a son, he'd look like Trayvon," and the fact that Martin was wearing a hoodie when he was shot. Then, around 19:18 EST (Fig. 9 (middle)), tweets about a newly released police surveillance video of Zimmerman start to emerge (shown in a few blue disks). The video showed that Zimmerman did not appear to be injured after the shooting, even though he claimed self-defense in shooting Martin. At the same time, tweets about the FoxNews comments slowed (pink region and disks south of the highlighted surveillance video tweet). Throughout the evening, tweets about newly released video increase and become an important part of the news landscape (blue disks and country at the top of Fig. 9 (bottom)).

TwitterScope has been online for a few months. During that time, the heaviest tweet rate encountered was about 34,000 tweets per hour for the term "news." The implementation of TwitterScope has been more than adequate to cope with this packet rate, while allowing responsive interaction.

## 6 CONCLUSION

We presented a general technique for visualizing real-time streams of text packets. It clusters packets into subtopics based on semantic analysis and, using a geographic map metaphor, renders subtopics as countries, with an attached textual precis. The approach relies on placement algorithms that promote the stability of the drawing to aid a user's comprehension. In particular, we introduced a new packing algorithm to address component stability.

We see two obvious algorithmic enhancements for our current approach, both dealing with visual stability. First, when we recreate the map, colors for countries are assigned without much consideration to stability. There has been recent work on achieving color assignment stability that we would like to incorporate [23]. Second, for certain topics, when we perform a periodic packing refresh, the map may change significantly. It would be good to avoid this discontinuity, either algorithmically or visually.

We would like to deploy TwitterScope as a visual search engine, allowing arbitrary keyword search instead of the fixed number of topics available in the current implementation. This will require a modified system architecture, including server-side query handling, but we don't envision any major changes to the algorithms described above.

At present, back-end processes handle most of the data filtering and analysis. As part of making the system more flexible, we expect to re-implement most of these processes in JavaScript so the computing and networking load can be shifted to the client side. This would mean shifting the clustering, mapping, dynamic layout, and component packing to the browser. While this is not easy to implement, having these algorithms available in clients would open new possibilities for other sophisticated visualization tasks.

## REFERENCES

[1] C. Albrecht-Buehler, B. Watson, and D. A. Shamma. Visualizing live text streams using motion and temporal pooling. *IEEE Computer Graphics and Applications*, 25(3):52–59, 2005.

[2] J. Alsakran, Y. Chen, D. Luo, Y. Zhao, J. Yang, W. Dou, and S. Liu. Real-time visualization of streaming text with a force-based dynamic system. *IEEE Computer Graphics and Applications*, 32(1):34–45, 2012.

[3] U. Brandes and S. R. Corman. Visual unrolling of network evolution and the analysis of dynamic discourse. In *IEEE INFOVIS'02*, pages 145–151, 2002.

[4] U. Brandes and M. Mader. A quantitative comparison of stress-minimization approaches for offline dynamic graph drawing. In M. J. van Kreveld and B. Speckmann, editors, *Graph Drawing*, volume 7034 of *Lecture Notes in Computer Science*, pages 99–110. Springer, 2011.

[5] U. Brandes and D. Wagner. A bayesian paradigm for dynamic graph layout. In G. D. Battista, editor, *Graph Drawing*, volume 1353 of *Lecture Notes in Computer Science*, pages 236–247. Springer, 1997.

[6] T. F. Cox and M. A. A. Cox. *Multidimensional Scaling*. Chapman and Hall/CRC, 2000.

[7] W. Cui, S. Liu, L. Tan, C. Shi, Y. Song, Z. Gao, H. Qu, and X. Tong. Textflow: Towards better understanding of evolving topics in text. *IEEE Trans. Vis. Comput. Graph.*, 17(12):2412–2421, 2011.

[8] W. Cui, Y. Wu, S. Liu, F. Wei, M. X. Zhou, and H. Qu. Context-preserving, dynamic word cloud visualization. *Computer Graphics and Applications*, 30:42–53, 2010.

[9] S. Diehl and C. Görg. Graphs, they are changing. In S. G. Kobourov and M. T. Goodrich, editors, *Graph Drawing*, volume 2528 of *Lecture Notes in Computer Science*, pages 23–30. Springer, 2002.

[10] T. Dwyer, K. Marriott, and P. J. Stuckey. Fast node overlap removal. In *Proc. 13th Intl. Symp. Graph Drawing (GD '05)*, volume 3843 of *LNCS*, pages 153–164. Springer, 2006.

[11] C. Erten, P. J. Harding, S. G. Kobourov, K. Wampler, and G. V. Yee. Graphael: Graph animations with evolving layouts. In G. Liotta, editor, *Graph Drawing*, volume 2912 of *Lecture Notes in Computer Science*, pages 98–110. Springer, 2003.

[12] S. I. Fabrikant, D. R. Montello, and D. M. Mark. The distance-similarity metaphor in region-display spatializations. *IEEE Computer Graphics & Application*, 26:34–44, 2006.

[13] S. Fortune. A sweepline algorithm for Voronoi diagrams. *Algorithmica*, 2:153–174, 1987.

[14] K. Freivalds, U. Dogrusöz, and P. Kikusts. Disconnected graph layout and the polyomino packing approach. In P. Mutzel, M. Jünger, and S. Leipert, editors, *Graph Drawing*, volume 2265 of *Lecture Notes in Computer Science*, pages 378–391. Springer, 2001.

[15] E. R. Gansner and Y. Hu. Efficient node overlap removal using a proximity stress model. In I. G. Tollis and M. Patrignani, editors, *Graph Drawing*, volume 5417 of *Lecture Notes in Computer Science*, pages 206–217. Springer, 2008.

[16] E. R. Gansner, Y. F. Hu, and S. G. Kobourov. Gmap: Visualizing graphs and clusters as maps. In *Proceedings of IEEE Pacific Visualization Symposium*, pages 201 – 208, 2010.

[17] E. R. Gansner, Y. Koren, and S. C. North. Graph drawing by stress majorization. In *Proc. 12th Intl. Symp. Graph Drawing (GD '04)*, volume 3383 of *LNCS*, pages 239–250. Springer, 2004.

[18] E. R. Gansner and S. North. An open graph visualization system and its applications to software engineering. *Software - Practice & Experience*, 30:1203–1233, 2000.

[19] E. R. Gansner and S. C. North. Improved force-directed layouts. In *Proc. 6th Intl. Symp. Graph Drawing (GD '98)*, volume 1547 of *LNCS*, pages 364–373. Springer, 1998.

[20] D. Goehlsdorf, M. Kaufmann, and M. Siebenhaller. Placing connected components of disconnected graphs. In S.-H. Hong and K.-L. Ma, editors, *APVIS: 6th International Asia-Pacific Symposium on Visualization 2007, Sydney, Australia, 5-7 February 2007*, pages 101–108. IEEE, 2007.

[21] B. Gretarsson, J. O'Donovan, S. Bostandjiev, T. Höllerer, A. U. Asuncion, D. Newman, and P. Smyth. Topicnets: Visual analysis of large text corpora with topic modeling. *ACM TIST*, 3(2):23, 2012.

[22] Herman, G. Melançon, and M. S. Marshall. Graph visualization and navigation in information visualization: A survey. *IEEE Transactions on Visualization and Computer Graphics*, 6(1):24–43, 2000.

[23] Y. Hu, S. Kobourov, and S. Veeramoni. Embedding, clustering and coloring for dynamic maps. In *Proceedings of IEEE Pacific Visualization Symposium*, 2012.

[24] O. Jin, N. N. Liu, K. Zhao, Y. Yu, and Q. Yang. Transferring topical knowledge from auxiliary long texts for short text clustering. In *Proceedings of the 20th ACM International Conference on Information and Knowledge Management*, CIKM '11, pages 775–784, New York, NY, USA, 2011. ACM.

[25] A. Marcus, M. S. Bernstein, O. Badar, D. R. Karger, S. Madden, and R. C. Miller. Twitinfo: aggregating and visualizing microblogs for event exploration. In D. S. Tan, S. Amershi, B. Begole, W. A. Kellogg, and M. Tungare, editors, *CHI*, pages 227–236. ACM, 2011.

[26] M. E. J. Newman. Modularity and community structure in networks. *Proc. Natl. Acad. Sci. USA*, 103:8577–8582, 2006.

[27] J. R. Shewchuk. Delaunay refinement algorithms for triangular mesh generation. *Computational Geometry: Theory and Applications*, 22:21–74, 2002.

[28] R. Sibson. Studies in the robustness of multidimensional scaling: Procrustes statistics. *Journal of the Royal Statistical Society, Series B (Methodological)*, 40:234–238, 1978.

[29] A. Silic and B. D. Basic. Visualization of text streams: A survey. In R. Setchi, I. Jordanov, R. J. Howlett, and L. C. Jain, editors, *KES (2)*, volume 6277 of *Lecture Notes in Computer Science*, pages 31–43. Springer, 2010.

[30] M. Tang and X. Yao. A memetic algorithm for VLSI floorplanning. *IEEE Transactions on Systems, Man, and Cybernetics, Part B*, 37(1):62–69, 2007.

[31] T. von Landesberger, M. Görner, and T. Schreck. Visual analysis of graphs with multiple connected components. In *Proceedings of the IEEE Symposium on Visual Analytics Science and Technology, IEEE VAST 2009, Atlantic City, New Jersey, USA, 11-16 October 2009*, pages 155–162, 2009.